\documentclass{llncs}
\usepackage{epsfig}

\begin{document}
\title{Multiscale modeling of biopolymer translocation 
through a nanopore}
\titlerunning{Multiscale modeling of biopolymer translocation}
\author{Maria Fyta\inst{1}
\and Simone Melchionna\inst{2} \and 
Efthimios Kaxiras\inst{1} \and Sauro Succi\inst{4}}
\institute{Department of Physics and Division of Engineering 
and Applied Sciences\\ Harvard University, Cambridge MA 02138, USA, \\
\email{mfyta@physics.harvard.edu, kaxiras@physics.harvard.edu}
\and INFM-SOFT, Department of Physics, Universit\`a di Roma 
{\it La Sapienza}\\
P.le A. Moro 2, 00185 Rome, Italy \\
\email{Simone.Melchionna@Roma1.infn.it}
\and Istituto Applicazioni Calcolo, CNR,  
Viale del Policlinico 137, 00161, Rome, Italy\\
\email{succi@iac.rm.cnr.it}
}

\maketitle

\begin{abstract}
We employ a multiscale approach to model the 
translocation of biopolymers through nanometer size pores.
Our computational scheme combines microscopic Langevin molecular 
dynamics (MD) with a mesoscopic lattice Boltzmann (LB) method 
for the solvent dynamics, explicitly taking into account the 
interactions of the molecule with the surrounding fluid.
Both dynamical and statistical aspects of the translocation 
process were investigated, by simulating polymers of 
various initial configurations and lengths. For a representative 
molecule size, we explore the effects of important parameters that 
enter in the simulation, paying particular attention to 
the strength of the molecule-solvent coupling and of the 
external electric field which drives the translocation process. 
Finally, we explore the connection between the generic polymers
modeled in the simulation and DNA, for which interesting recent 
experimental results are available. 
\end{abstract}


\section{Introduction
\label{sec_intro}
}

Biological systems exhibit a complexity and diversity far richer
than the simple solid or fluid systems traditionally studied in 
physics or chemistry. The powerful quantitative methods developed 
in the latter two disciplines to analyze the behavior of 
prototypical simple systems are often difficult to extend to the 
domain of biological systems. Advances in computer technology 
and breakthroughs in simulational methods have been constantly 
reducing the gap between quantitative models and actual biological 
behavior. The main challenge remains the wide and disparate range
of spatio-temporal scales involved in the dynamical evolution of
complex biological systems. In response to this challenge, 
various strategies have been developed recently, 
which are in general referred to as ``multiscale modeling''. 
These methods are based on composite computational schemes
in which information is exchanged between the scales.

We have recently developed a multiscale framework which 
is well suited to address a class of biologically related problems.
This method involves different levels 
of the statistical description of matter (continuum and 
atomistic) and is able to handle different scales through 
the spatial and temporal coupling of a {\it mesoscopic} fluid
solvent, using the lattice Boltzmann method \cite{LBE} (LB), with the 
atomistic level, which employs explicit molecular dynamics (MD).
The solvent dynamics does not require any form of statistical 
ensemble averaging as it is represented through a discrete set 
of pre-averaged probability distribution functions,
which are propagated along straight particle trajectories.
This dual field/particle nature greatly facilitates the coupling 
between the mesoscopic fluid and the atomistic level, which 
proceeds seamlessy in time and only requires standard 
interpolation/extrapolation for information-transfer in physical 
space. Full details on this scheme are reported in 
Ref. \cite{ourLBM}.
We must note that to the best of our knowledge, 
although LB and MD with Langevin dynamics 
have been coupled before \cite{DUN}, 
this is the first time that such a coupling is put in 
place for long molecules of biological interest.

Motivated by recent experimental studies, we apply
this multiscale approach to the translocation of a 
biopolymer through a narrow pore. These kind
of biophysical processes are important in phenomena like 
viral infection by phages, inter-bacterial DNA transduction or 
gene therapy \cite{TRANSL}. In addition,
they are believed to open a way for ultrafast DNA-sequencing 
by reading the base sequence as the biopolymer passes through 
a nanopore.
Experimentally, translocation is observed {\it in vitro}
by pulling DNA molecules through micro-fabricated solid state or
membrane channels under the effect of a localized electric
field \cite{EXPRM}. From a theoretical point of view, 
simplified schemes \cite{statisTrans} and non-hydrodynamic 
coarse-grained or microscopic models \cite{DynamPRL,Nelson} 
are able to analyze universal features of the translocation 
process. This, though, is a complex phenomenon involving the 
competition between many-body interactions at the atomic 
or molecular scale, fluid-atom hydrodynamic coupling, as well 
as the interaction of the biopolymer with wall molecules in 
the region of the pore. A quantitative description of this 
complex phenomenon calls for state-of-the art modeling, towards which
the results presented here are directed.

\section{Numerical Set-up
\label{sec_numer}
}

In our simulations we use a three-dimensional box 
of size $N_x \times N_x/2 \times N_x/2 $ in units of the 
lattice spacing $\Delta x$. The box contains both the polymer 
and the fluid solvent. The former is initialized via a 
standard self-avoiding random walk algorithm and further relaxed 
to equilibrium by Molecular Dynamics. The solvent is 
initialized with the equilibrium distribution corresponding to a
constant density and zero macroscopic speed. Periodicity 
is imposed for both the fluid and the polymer in all directions.
A separating wall is located in the mid-section of the $x$ 
direction, at $x/\Delta x= N_x/2$, with a square hole of 
side $h=3 \Delta x$ at the center, through which the polymer 
can translocate from one chamber to the other. For polymers 
with up to $N=400$ beads we use $N_x = 80$; for larger 
polymers $N_x = 100$. At $t=0$ the polymer resides entirely 
in the right chamber at $x/\Delta x>  N_x/2$. 
The polymer is advanced in time according to the following
set of Molecular Dynamics-Langevin equations for the bead positions 
$\vec{r}_p$ and velocities $\vec{v}_p$ (index $p$ runs over all beads):
\begin{eqnarray}
\label{MD}
M_p \frac{d \vec{v}_p}{dt} &=&
-\sum_q \partial_{\vec{r}_p} V_{LJ}(\vec{r}_p-\vec{r}_q)+
 \gamma (\vec{u}_p-\vec{v}_p)+
 M_p \vec{\xi}_p
-\lambda_p \partial_{\vec{r}_p} \kappa_p
\end{eqnarray}
These interact among themselves through 
a Lennard-Jones potential with $\sigma=1.8 $ and $\varepsilon= 10^{-4}$:
\begin{equation}
V_{LJ}(r) = 4 \varepsilon \left[ \left(\frac{\sigma}{r}\right)^{12} 
- \left(\frac{\sigma}{r}\right)^{6}\right]
\end{equation}
This potential is augmented by an angular harmonic term to account 
for distortions of the angle between consecutive bonds.
The second term in Eq.(\ref{MD}) represents the 
mechanical friction between a bead and the surrounding fluid, 
$\vec{u}_p$ is the fluid velocity evaluated at the bead position
and $\gamma$ the friction coefficient.
In addition to mechanical drag, the polymer feels the effects 
of stochastic fluctuations of the fluid environment, through the random 
term, $\vec{\xi}_p$. This is related to the third term in Eq.(\ref{MD}),
which is an incorrelated random term with zero mean. Finally,
the last term in Eq.(\ref{MD}) is the reaction force resulting 
from $N-1$ holonomic constraints for molecules modelled with rigid 
covalent bonds. The bond length is set at $b=1.2$ and
$M_p$ is the bead mass equal to 1.

\begin{figure}
\begin{center}
\epsfig{file=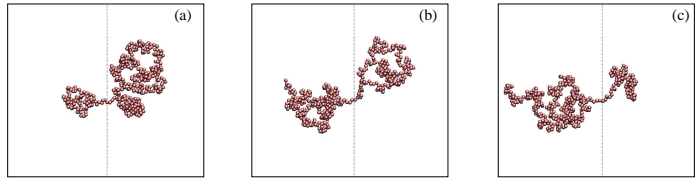,width=0.75\textwidth}
\caption{Snapshots of a typical event: a polymer 
($N=300$) translocating
from the right to the left is depicted at a time
equal to (a) 0.11, (b) 0.47, and (c) 0.81 of the
total time for this translocation.
The vertical line in the middle of each panel shows the wall.}
\label{FIG1}
\end{center}
\end{figure}

Translocation is induced by a constant electric force ($F_{drive}$) 
which acts along the $x$ direction and is confined in a rectangular
channel of size $3\Delta x \times \Delta x \times \Delta x$ 
along the streamwise ($x$ direction) and cross-flow 
($y,z$ directions). 
The solvent density and kinematic viscosity are $1$ and
$0.1$, respectively, and the temperature is $k_B T=10^{-4}$.
All parameters are in units of the LB timestep $ \Delta t$ 
and lattice spacing $\Delta x$, which we set equal to 1.
Additional details have been presented in Ref. \cite{ourLBM}.
In our simulations we use $F_{drive} =0.02$ and a friction 
coefficient $\gamma=0.1$. It should be kept in mind that 
$\gamma$ is a parameter governing both the structural relation of 
the polymer towards equilibrium and the strength of the coupling 
with the surrounding fluid. The MD timestep is a fraction of the 
timestep for the LB part $\Delta t=m\Delta t_{MD}$, 
where $m$ is a constant typically set at $m=5$.
With this parametrization, the process falls in the fast 
translocation regime, where the total
translocation time is much smaller than the Zimm relaxation time.
We refer to this set of parameters as our ``reference''; 
we explore the effect of the most important parameters for certain 
representative cases.

\section{Translocation dynamics
\label{sec_trans}
}

Extensive simulations of a large number of translocation events 
over $100-1000$ initial polymer configurations for each
length confirm that most of the time during the
translocation process the polymer assumes the form of two 
almost compact blobs on either side of the wall: one of them
(the untranslocated part, denoted by $U$) is contracting and the other 
(the translocated part, denoted by $T$) 
is expanding. Snapshots of a typical translocation event 
shown in Fig.~\ref{FIG1} strongly support this picture. 
A radius of gyration $R_I(t)$ (with $I=U,T$) is assigned to each of 
these blobs, following a static scaling 
law with the number of beads $ N_{I}$:
$R_I(t) \sim N_{I}^{\nu}(t)$ with $\nu \simeq 0.6$ being 
the Flory exponent for a three-dimensional self-avoiding 
random walk. Based on the conservation of polymer length, 
$N_U+N_T = N_{tot}$, 
an effective translocation radius can be defined as 
$R_E(t) \equiv (R_T(t)^{1/\nu} + R_U(t)^{1/\nu})^{\nu}$.
We have shown that $R_E(t)$
is approximately constant for all times when the static scaling 
applies, which is the case throughout the process
except near the end points (initiation
and completion of the event) \cite{ourLBM}. 
At these end points, deviations from the mean field picture, 
where the polymer is represented as two uncorrelated compact 
blobs, occur. The volume of the polymer also changes after 
its passage through the pore.
At the end, the radius of gyration is considerably smaller than it
was initially: $R_T(t_X)<R_U(0)$, where $t_X$ is the total
translocation time for an individual event. For our
reference simulation an average over a few hundreds of events 
for $N=200$ beads showed that
$\lambda_R=R_T(t_X)/R_U(0)\sim 0.7$. This reveals the 
fact that as the polymer passes through the pore 
it is more compact than it was at the initial stage 
of the event, due to incomplete relaxation.

\begin{figure}
\begin{center}
\includegraphics[width=0.75\textwidth]{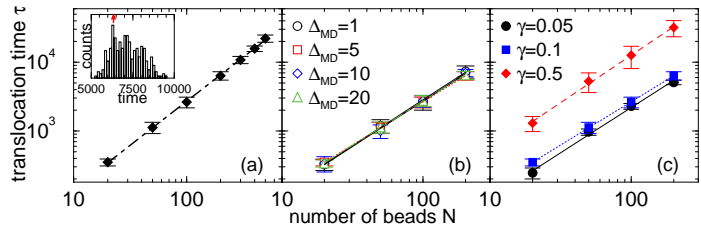}
\caption{(a) Scaling of $\tau$ with the number of beads $N$. 
Inset: distribution of translocation times over 300 events for
$N=200$. Time is given in units of the LB timestep. The 
arrow shows the most probable translocation time for this length.
Effect of the various parameters on the scaling law: 
(b) changing the value of the MD timestep  
($\Delta t_{MD}$); (c) changing the value of the 
solvent-molecule coupling coefficient $\gamma$.}
\label{FIG2}
\end{center}
\end{figure}

The variety of different initial polymer realizations 
produce a scaling law dependence of the translocation times
on length \cite{Nelson}. By accumulating all events for each length, 
duration histograms were constructed. The resulting 
distributions deviate from simple gaussians and are 
skewed towards longer times (see Fig.~\ref{FIG2}(a) inset). Hence, 
the translocation time for each length is not assigned to 
the mean, but to the most probable time ($t_{max}$), 
which is the position of the maximum in the histogram
(noted by the arrow in the inset of Fig.~\ref{FIG2}(a) for
the case $N=200$). By calculating the most probable times
for each length, a superlinear relation between the
translocation time $\tau$ and the number of beads $N$ 
is obtained and is reported in Fig.~\ref{FIG2}(a). The exponent
in the scaling law $\tau (N) \sim N^{\alpha}$ is calculated as
$\alpha \sim 1.28\pm0.01$, for lengths up to $N=500$ beads.
The observed exponent is in very good agreement 
with a recent experiment on double-stranded DNA
translocation, that reported $\alpha\simeq1.27\pm0.03$ \cite{NANO}. 
This agreement makes it plausible that 
the generic polymers modeled in our simulations can be 
thought of as DNA molecules; we return to this issue 
in section \ref{sec_realDNA}.

\section{Effects of parameter values
\label{sec_param}
}

We next investigate the effect that the various 
parameters have on the simulations, using as standard of 
comparison the parameter set that we called the ``reference'' case. 
For all lengths and parameters about
100 different initial configurations were generated
to assess the statistical and dynamical features of the 
translocation process.
As a first step we simulate polymers of different lengths
($N=20-200$). Following a procedure similar to the previous
section we extract the scaling laws for the translocation
time and their vatiation with the friction coefficient
$\gamma$ and the MD timestep $\Delta t_{MD}$. The
results are shown in Fig.~\ref{FIG2}(b) and (c). 
In these calculations the error bars were also taken into 
account. The scaling exponent for our reference
simulation ($\gamma=0.1$) presented in
Fig.~\ref{FIG2}(a) is $\alpha\simeq1.27\pm0.01$ when
only the lengths up to $N=200$ are included.
The exponent for smaller damping 
($\gamma=0.05$) is $\alpha\simeq1.32\pm0.06$, and for larger
($\gamma=0.5$) $\alpha\simeq1.38\pm0.04$. 
By increasing $\gamma$ by one order of magnitude the time
scale rises by approximately one order of magnitude, showing
an almost linear dependence of the translocation time
with hydrodynamic friction; we discuss this
further in the next section. However, for larger $\gamma$, thus
overdamped dynamics and smaller influence of the driving force, 
the deviation from the 
$\alpha=1.28$ exponent suggests a systematic departure from
the fast translocation regime. 
Similar analysis for various values of $\Delta t_{MD}$ 
shows that the exponent becomes $\alpha\simeq1.34\pm0.04$
when $\Delta t_{MD}$ is equal to the LB timestep ($m=1$); for
$m=10$ the exponent is $\alpha\simeq1.32\pm0.04$, while
for $m=20$, $\alpha\simeq1.28\pm0.01$
with similar prefactors.

\begin{figure}
\begin{center}
\epsfig{file=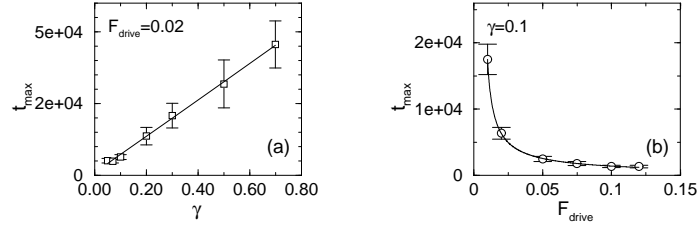,width=0.75\textwidth}
\caption{Variation of $t_{max}$ with
(a) $\gamma$, and (b) $F_{drive}$ for $N=200$ beads.}
\label{FIG3}
\end{center}
\end{figure}

\begin{figure}
\begin{center}
\epsfig{file=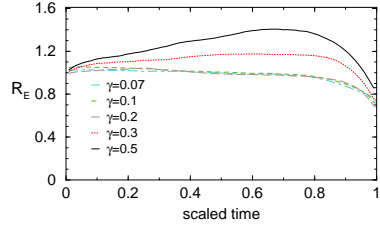,width=0.4\textwidth}
\caption{The dependence of the effective radii of gyration $R_E(t)$ 
on $\gamma$ ($N=200$). Time and $R_E$ are scaled with 
respect to the total translocation time and $R_U(0)$ for each case.}
\label{FIG4}
\end{center}
\end{figure}

We next consider what happens when we fix the length to $N=200$ and
vary $\gamma$ and the pulling force $F_{drive}$. For 
all forces used, the process falls in the fast translocation regime.
The most probable time ($t_{max}$) for each 
case was calculated and the results are shown in Fig.~\ref{FIG3}.
The dependence of $t_{max}$ on $\gamma$ is linear related to the 
linear dependence of $\tau$ on $\gamma$, mentioned in the previous section.
The variation of $t_{max}$ with $F_{drive}$ follows an inverse
power law: $t_{max}\sim 1/F_{drive}^{\mu}$, with $\mu$ of the order 1.
The effect of $\gamma$ is further explored in relation to the 
effective radii of gyration $R_E$, and is presented in Fig.~\ref{FIG4}.
The latter must be constant when the static scaling
$R\sim N^{0.6}$ holds. This is confirmed for small $\gamma$
up to about $0.2$. As $\gamma$ increases, $R_E$ is no more 
constant with time, and shows interesting behavior: 
it increases continuously up to a point where a large fraction of 
the chain has passed through the pore and subsequently drops  
to a value smaller than the initial $R_U(0)$. 
Hence, as $\gamma$ increases large deviations from the static
scaling occur and the translocating polymer can no longer 
be represented as two distinct blobs. In all
cases, the translocated blob becomes more compact.
For all values of $\gamma$ considered,
$\lambda_R$ is always less than unity ranging 
from $0.7$ ($\gamma$=0.1) to $0.9$ ($\gamma$=0.5) 
following no specific trend with $\gamma$.

\section{Mapping to real biopolymers
\label{sec_realDNA}
}

As a final step towards connecting our computer simulations
to real experiments and after having established the agreement 
in terms of the scaling behavior, we investigate the mapping 
issue of the polymer beads to double-stranded DNA.
In order to interpret our results in terms of physical units,
we turn to the persistence length ($l_p$) of the semiflexible 
polymers used in our simulations. Accordingly, we use the formula 
for the fixed-bond-angle model of a worm-like chain \cite{lpWCL}: 
\begin{equation}
l_p=\frac{b}{1-\cos\langle \theta \rangle} 
\end{equation}
where $\langle\theta\rangle$ is complementary to the average 
bond angle between adjacent bonds. In lattice units 
($\Delta x$) an average persistence length for the polymers
considered, was found to be approximately $12$. 
For $\lambda$-phage DNA $l_p\sim 50$ nm \cite{lpDNA}
which is set equal to $l_p$ for our polymers. Thereby,
the lattice spacing is $\Delta x \sim 4$ nm, which is
also the size of one bead. Given that the base-pair 
spacing is $\sim0.34$ nm, one bead maps
approximately to $12$ base pairs. With this mapping, the pore size is
about $\sim12$ nm, close to the experimental pores which are
of the order of $10$ nm. The polymers presented here correspond to
DNA lengths in the range $0.2-6$ kbp. The DNA lengths used
in the experiments are larger (up to $\sim$ 100kbp);
the current multiscale approach can be extended to handle
these lengths, assuming that appropriate computational
resources are available.

\begin{figure}
\begin{center}
\epsfig{file=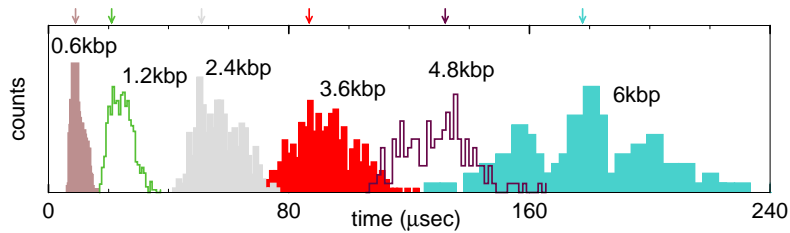,width=0.85\textwidth}
\caption{Histograms of calculated translocation times for a 
large number of events and different DNA lengths.
The arrows link to the most probable time ($t_{max}$) for each case.}
\label{FIG5}
\end{center}
\end{figure}

Choosing polymer lengths that match experimental data 
we compare the corresponding experimental duration histograms 
(see Fig.~1c of Ref.~\cite{NANO}) to the theoretical ones. This
comparison sets the LB timestep to $\Delta t\sim8$ nsec.
In Fig.~\ref{FIG5} the time distributions for representative
DNA lengths simulated here are shown. In this figure, 
physical units are used
according to the mapping described above and promote comparison
with similar experimental data \cite{NANO}.
The MD timestep for $m=5$ will then be $t_{MD}\sim40$ nsec
indicating that the MD timescale related to the coarse-grained 
model that handles the DNA molecules is significantly stretched 
over the physical process. Exact match to all the experimental 
parameters is of course not feasible with coarse-grained
simulations. However, essential features of DNA translocation 
are captured, allowing the use of the current approach to model similar
biophysical processes that involve biopolymers in solution.
This can become more efficient by exploiting the freedom of 
further fine-tuning the parameters used
in this  multiscale model.

\section{Conclusions
\label{sec_concl}
}

In summary, we applied a multiscale methodology to 
model the translocation of a biopolymer through 
a nanopore. Hydrodynamic correlations between the
polymer and the surrounding fluid have explicitly been 
included. The polymer obeys a static scaling except near 
the end points for each event (initiation and 
completion of the process) and the translocation 
times vary exponentially with the polymer length.
A preliminary exploration of the effects of the most important  
parameters used in our simulations was also presented,
specifically the values of the friction coefficient and the pulling
force describing the effect of the external electric 
field that drives the translocation. 
These were found to significantly affect the dynamic features
of the process. Finally, our generic polymer models were 
directly mapped to double-stranded DNA and a comparison to 
experimental results was discussed. 

\subsubsection*{Acknowledgments.} 
MF acknowledges support by Harvard's Nanoscale Science and 
Engineering Center, funded by NSF (Award No. PHY-0117795).

\end{document}